\documentclass{article}
\usepackage{spconf,amsmath,graphicx}

\def\isconfidential{1}
\def\includecomments{0}

\usepackage{sty/commands}
\usepackage{euscript}

\newcommand{\ray}[1]{\if\includecomments1\textcolor{blue}{[Ray:~#1]}\fi}
\newcommand{\mg}[1]{\if\includecomments1\textcolor{green!50!black}{[MG:~#1]}\fi}

\title{Semantic Property Maps for Driving Applications}

\name{Marcus Greiff, Ray Zhang, Takeru Shirasawa, John Subosits}
\address{Toyota Research Institute, 4440 El Camino Real, Los Altos, USA}

\begin{document}

\maketitle

\begin{abstract}

We consider the problem of estimating the parameters of a vehicle dynamics model for predictive control in driving applications. Instead of solely using the instantaneous parameters estimated from the vehicle signals, we combine this with cameras and update a probabilistic map with parameter estimates and semantic information using Bayesian moment matching. Key to this approach is the map representation, which is constructed with conjugate priors to the measurement likelihoods and defined in the same path coordinates as the vehicle controller, such that the map can be externalized to provide a local representation of the parameter likelihoods that vary in space. The result is a spatial map of vehicle parameters adapted online to enhance the driving control system. We provide theoretical guarantees on the smoothness of relevant parameter likelihood statistics as a function of space, which is critical for their use in predictive control.
\end{abstract}
\begin{keywords}
Bayesian Inference, Moment Matching, Online Learning, Semantic Mapping, Parameter Estimation. \emph{This work was presented at the ISPAV workshop at ICIP 2025.}
\end{keywords}
\section{Introduction}\label{sec:intro}
Accurately representing an uncertain world is critical to ensuring safety in modern advanced driving assistance systems (ADAS) and autonomous driving (AD). This requires exteroceptive sensing, often using LiDARs, radars, and cameras. Depending on the application, the resulting output is typically a semantic map of obstacles to be avoided or geometric representations, which inform safety systems and often explicitly constrain the planning and control algorithms~\cite{mukhtar2015vehicle}. 

Semantic mapping is often used to construct a model based on multiple observations that is both spatially and semantically consistent~\cite{nuchter2008towards}. Classical discrete semantic mapping techniques combine 2D semantic predictions from machine learning algorithms with geometric representations in 3D, such as voxels or surfaces, then fuse the semantics via moving averages or conditional random fields (CRFs)~\cite{mccormac2017semanticfusion}.  In contrast, continuous mapping techniques use Gaussian processes (GPs) and their efficient approximations as the map representations, and can produce continuous occupancy updates~\cite{ o2012gaussian,  Doherty2019}, while semantics are integrated via Bayesian inference~\cite {gan2020bayesian, wilson2024convbki}. With recent advances in neural fields, fully differentiable scene representations can integrate semantic feature embeddings and support continuous semantic queries~\cite{ zhou2024feature, wilson2025latentbki}. However, these methods primarily operate with Cartesian coordinates, which complicates integration control and planning algorithms that operate in path coordinates.

In addition to semantics, driving applications may require richer map representations. Vehicle parameters such as friction coefficients are spatially correlated and crucial to safe autonomous driving and navigation in challenging scenarios, including winter driving and racing~\cite{svensson2021traction, thompson2024adaptive}. Early works on estimating such properties include model-based estimation of vehicle and tire models~\cite{gustafsson1997slip} and offline supervised learning~\cite{panahandeh17, du20} from proprioceptive sensing. Online learning of properties is often employed to adapt to changes in terrain conditions and vehicle models~\cite{dallas2020online, werner2025gripmap}. Recent advances in multimodal learning enable fusion of visual and physical properties via GP regression~\cite{ svensson2021fusion, margolis2023learning}. To impose additional structure in the latent space, the recent work~\cite{ewen2022these,ewen2024you} models the property as a Gaussian mixture whose weights depend on visual segmentation priors in a Bayesian framework.

Inspired by the methods in~\cite{ewen2022these,ewen2024you} for legged locomotion, we hypothesize a correlation between semantic meanings and the physical properties of the vehicle in a map, and estimate this directly~\cite{du20}. However, in contrast to these works, we seek a representation that is more suitable for driving. Specifically, we require that (i) the moments of the vehicle model parameters are continuously differentiable in the position of the vehicle; (ii) inference is done in {path coordinates} relative to a known road geometry (see e.g.,~\cite{Goh2019}); and (iii) the map posterior can be externalized and used directly in the optimization problems of the controller and planner~\cite{thompson2024adaptive, lew2024risk}.

\begin{figure*}[t!]
    \centering
    \begin{tikzpicture}
\node[inner sep=0pt] (russell) at (0,0)
    {\includegraphics[width=\textwidth]{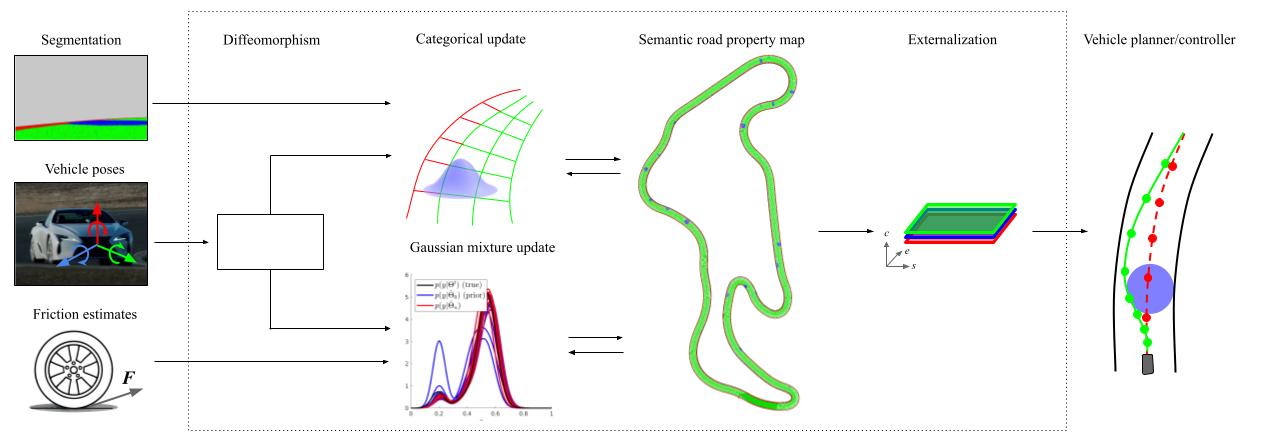}};
    \node[inner sep=0pt] at (-5.09,-0.3) {\scalebox{.55}{$\phi:\Real^2_{\mathrm{BEV}}\mapsto \Real^2_{\mathrm{SE}}$}};
    
    \node[inner sep=0pt] at (-0.6,1.05) {\scalebox{.55}{$\Pcal_{k+1}$}};
    \node[inner sep=0pt] at (-0.6,0.50) {\scalebox{.55}{$\Pcal_{k}$}};

    \node[inner sep=0pt] at (-0.6,-1.47) {\scalebox{.55}{$\Pcal_{k+1}$}};
    \node[inner sep=0pt] at (-0.6,-2.02) {\scalebox{.55}{$\Pcal_{k}$}};
    
    \node[inner sep=0pt] at (4.6,-2.8) {\scalebox{.7}{{\color{white!50!black}Scope of paper}}};
    
    \end{tikzpicture}
    \vspace{-25pt}
    
    \caption{The semantic property map performs Bayesian updates using visual information and vehicle parameter estimates.}
    \label{fig:method}
\end{figure*}
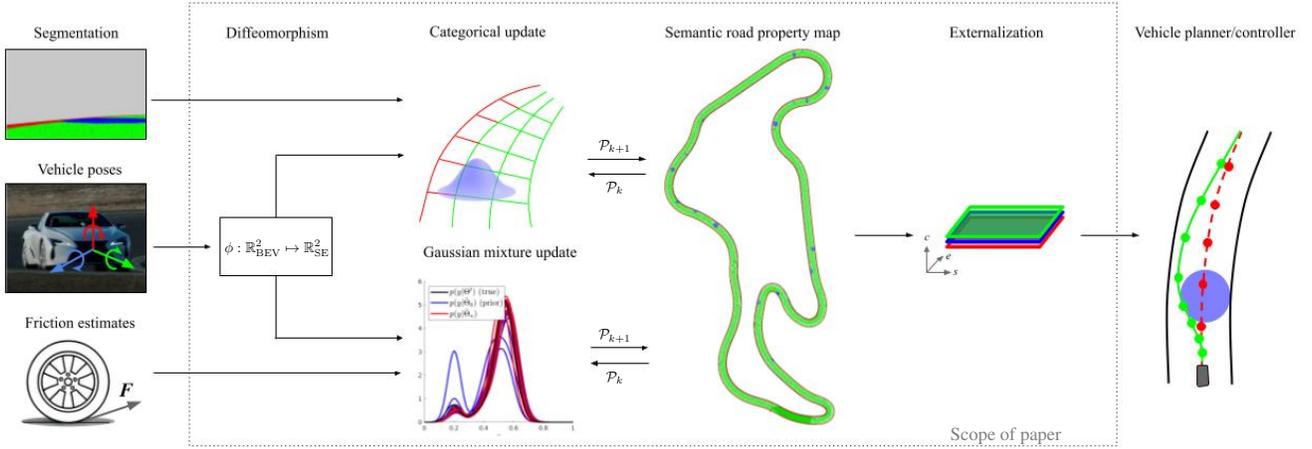

\subsection{Contributions}

We extend the voxel-based probabilistic map in~\cite{ewen2024you} to a generic map in a road's path coordinates~\cite{Goh2019}, and demonstrate improved predictions of friction coefficients \emph{ahead} of the vehicle (see Fig.~\ref{fig:method}). By interpolating Gaussian mixture and categorical measurement likelihoods with sparse kernels, the map posteriors can be computed efficiently using Bayesian moment matching. Additionally, we show that this update is local and that relevant moments of the posterior property likelihoods are continuously differentiable in space.

\section{Preliminaries}\label{sec:related}
 The usual $K$-dimensional Dirichlet distribution is defined with a density $\mathrm{D}(\wvec;\avec) \propto \prod_{i=1}^K w_i^{a_i-1}$, and the categorical distribution is defined with a point mass function $\mathrm{C}(y;\wvec)\propto\prod_{i=1}^K w_i^{[y=i]}$, where $[\cdot]$ denotes the Iverson bracket, that is, $[y=i]=1$ if $y=i$ and 0 otherwise. If given an integer argument, $[K] = \{1,...,K\}\subset \mathbb{N}$. We consider normal-gamma distributions with PDF $\mathrm{N}\Gamma(m, \tau; \mu, \lambda, \alpha, \beta)\propto \tau^{\alpha-\frac{1}{2}}\mathrm{exp}(-\beta\tau -\frac{\lambda\tau(x - \mu)^2}{2})$. The 2D Bird's-Eye View (BEV) Cartesian and path coordinates are expressed in $\mathbb{R}^2_{\mathrm{BEV}}$ and $\mathbb{R}^2_{\mathrm{SE}}$, respectively. The notation $f\in \Ccal^n(A, B)$ denotes a $n$-times continuously differentiable (CD) function from $A$ to $B$. 
 
 A \emph{semantic property map} (SPM) is a function $f\in \Ccal^{n}(\Real^{2}_{\mathrm{SE}}, \Real) $ that relates a spatial location in path coordinates to a property using a latent semantic representation. We associate the property map with a distribution defined by a PDF $p(\Theta)$, where $\Theta$ is the state of the map. Given a measurement $y$, we compute the posterior using Bayes' rule
\begin{equation}\label{eq:bayes_update}
p(\Theta|y) = \frac{p(y|\Theta)p(\Theta)}{p(y)},
\end{equation}
which is often simplified by choosing an appropriate (conjugate) prior.\footnote{For example, the conjugate prior of the likelihood $C(c; \wvec)$ is $D(\wvec; \avec)$.} A prior $p(\Theta)$ belonging to a parametric family of distributions $\mathbb{P}$ is said to be conjugate for the likelihood $p(y|\Theta)$ if and only if the posterior $p(\Theta|y)$ belongs to $\mathbb{P}$. If priors are not conjugate, this generally leads to an exponential increase in the number of parameters required to represent the posterior exactly. For tractable implementation in such cases, we need a method of projecting the posterior onto $\mathbb{P}$, which can be done using Bayesian Moment Matching (BMM)~\cite{jaini2016online}. Here, we find the sufficient statistics of the densities in $\mathbb{P}$, denoted $g\in \mathbb{S}$, and equate these with the moments of the posterior $p(\Theta|y)$ in expectation\footnote{For the Dirichlet distribution, $g=\{(w_i,w_i^2)|i = 1,...,K\}$ defines $\mathbb{S}$.}. That is, for some new density $\bar{p}\in\mathbb{P}$, we solve $\mathbb{E}_{\bar{p}(\Theta)}[g] = \mathbb{E}_{p(\Theta|y)}[g]$ for the parameters of $\bar{p}$ to compute an approximate $\bar{p}(\Theta)\simeq p(\Theta|y)$. While approximate, such methods have proven useful in machine learning, filtering, and semantic mapping for robotic locomotion~\cite{ewen2024you}.

\section{Semantic Property Maps}\label{sec:maps}
Inspired by the map representations in \cite{gan2020bayesian, ewen2024you}, we model the semantics from a segmentation model with a closed set of $K$ classes using categorical likelihoods,  and the property estimates (computed by parameter estimators) with Gaussian mixture likelihoods. For the driving application, we seek a representation that can accommodate (i) continuously differentiable moments of the measurement likelihoods in space and (ii) facilitate updates with both semantic and Gaussian mixture likelihoods. Informed by the conjugate priors to these likelihoods, we construct the map using independent Dirichlet-normal-gamma distributions defined by a PDF
\begin{equation}
p(\Theta; \Pcal) = \prod_{\ell=1}^{L}\mathrm{D}(\wvec^{\ell}|\avec^{\ell})\prod_{i=1}^K\mathrm{N}\Gamma(m_i, \tau_i| \mu_i, \lambda_i, \alpha_i,\beta_i),\label{eq:priormodel}
\end{equation}
in $\Theta = \{\{\wvec^{\ell}\in\Real^{K}_{>0}\}_{\ell=1}^L, \{(m_i,\tau_i)\in\Real\times \Real_{>0}\}_{i=1}^K\}$ with
\begin{equation}
    \Pcal =
    \begin{Bmatrix}
    \{\avec^{\ell}\in\Real_{>0}^K\;:\;\ell\in[L]\},\\
    \{(\mu_i, \lambda_i, \alpha_i,\beta_i)\in\Real\times \Real_{\geq 0}^3\;:\;i \in[K]\}
    \end{Bmatrix},
\end{equation}
where $\Theta$ is the state of the map and $\Pcal$ denote its parameters. For this map representation, its sufficient moments are $g = \{(w_i^{\ell}, (w_i^{\ell})^2)|\ell\hspace{-1pt}\in\hspace{-1pt}[L], i\hspace{-1pt}\in\hspace{-1pt}[K]\}\hspace{-1pt}\cup\hspace{-1pt} \{(m_i,\tau_i,\tau_i^2, m_i^2\tau_i)|i\hspace{-1pt}\in\hspace{-1pt}[K]\}$. Each Dirichlet component is associated with a position in a set $\{\vvec_{\ell}\in\Real_{\mathrm{SE}}^2\}_{\ell = 1}^L$. To facilitate vehicle control, we define these points in \emph{path coordinates} using an efficient diffeomorphism defined with splines, which permits measurements in the image space to be related to specific measurement likelihoods expressed in $\Pcal$. Additionally, we consider continuous property likelihoods defined by interpolation, such that relevant moments of these measurement likelihoods are \emph{continuous in the path coordinates}, thus providing well-defined gradients for predictive control (see Fig.~\ref{fig:method}). This is in contrast to maps that model properties as constant over grid cells, as in~\cite{ewen2024you}.

\subsection{Projections and Diffeomorphisms}\label{sec:map:ipm}

Let $\ubar{c}\in[K]$ be a class sampled at pixels $(u,v)$ in an image, as generated by the segmentation model (see Fig.~\ref{fig:method}). Assuming the vehicle is moving in a plane, we relate this measurement to a likelihood expressed in $\Pcal$ with an inverse perspective mapping (IPM). This projects the pixel coordinates $(u,v)$ to a point on a plane defined by a normal $\nvec\in\Real^3$ and passing through a point $\rvec\in\Real^3$. With a camera matrix $\K\in\Real^{3\times 3}$, we can express a ray associated with the pixel $(u,v)$ as $\lambda \K^{-1}(u; v; 1)$, where $\lambda\in\Real$. If $\T\in\mathsf{SE(3)}$\footnote{With a slight abuse of notatoin, we let $\avec,\bvec\in\Real^3$ and permit the notation $\avec = \T\bvec$, which is more properly written as $\avec = \begin{bmatrix}\I_3 & \Z\end{bmatrix}\T(\bvec^{\top}, 1)^{\top}
$. } is a transform from the global frame to the camera frame, we obtain the distance to the point of intersection as
\begin{subequations}
\begin{equation}
\lambda^{\star} = \frac{\nvec\cdot \rvec}{\nvec \cdot \T\K^{-1}(u, v, 1)^{\top}},
\end{equation}
with a corresponding intersection point
\begin{equation}
\xvec^{\mathrm{3D}} = \lambda^{\star}\T\K^{-1}(u, v, 1)^{\top}\in\Real^3.
\end{equation}
\end{subequations}
For simplicity, we let $\nvec=(0;0;1)$ and $\rvec = (0,0,z)$ where $z$ is the elevation of the ground plane, and represent this in a Birds-eye View (BEV) perspective as $(x, y)\in\Real^{2}_{\mathrm{BEV}}\subset\Real^2$. While inference can be done in these coordinates~\cite{ewen2024you}, vehicle control problems are often posed in path coordinates~\cite{thompson2024adaptive}. We therefore let $(s,e)\in[s_{\min},s_{\max}]\times [e_{\min},e_{\max}]=\Real^2_{\mathrm{SE}}$ denote path coordinates, where $s$ is a distance traveled along a path $\gvec$, and $e$ is a signed distance in the normal direction of $\gvec$. The function $\gvec$ can represent both open and closed paths in the plane. In the latter, we let $|s-s^{\prime}|=\mathrm{argmin}_{n\in\mathbb{Z}}(s-s^{\prime}+n(s_{\max}- s_{\min}))$. With these definitions, we express a diffeomorphism $\phi\;:\;\Real^2_{\mathrm{BEV}}\mapsto \Real^2_{\mathrm{SE}}$ and its inverse as follows.

\begin{lemma}\label{lem:diffeomorphism}
If $\gvec\in\mathcal{C}^2([s_{\min},s_{\max}],\Real^2_{\mathrm{BEV}})$ with maximum curvature less than $e_{\max}^{-1}$ and $\min_{|s-s^{\prime}|> \pi e_{\max}}\|\gvec(s)- \gvec(s^{\prime})\|_2 > 2 e_{\max}$, then a diffeomorphism $\phi$ exists with
\begin{equation}
\hspace*{-4pt}\begin{cases}\label{eq:optprob}
s = \mathrm{argmin}_{s^{\prime}} \|\gvec(s^{\prime}) - (x,y)\|_2^2\\
e = \mathrm{sign}(\gvec^{\prime}(s)\times (\gvec(s) - (x,y)))\|\gvec(s^{\prime}) - (x,y)\|_2
\end{cases}\hspace*{-4pt}
\end{equation}
and $(x,y) = ([\gvec(s)]_1 + e[\gvec^{\prime}(s)]_2, [\gvec(s)]_2 - e[\gvec^{\prime}(s)]_1)$.
\end{lemma}
In this work, we regress $\gvec$ to known road geometries by solving a constrained least-squares problem over a sequence of $M$ B\'ezier curves and enforcing continuity in the first two derivatives at the end-points. In practice, it is possible to estimate these parameters online, if posed as a filtering problem~\cite{berntorp2024framework}. In the following, we regress the geometry of a 2-mile-long road using $M=40$ splines of degree 6, as shown in Fig.~\ref{fig:method}.

\subsection{Sparse Kernels and Interpolated Likelihoods}
\ray{We probably need to draw the boundaries between this work and some GP/RKHS-based continuous mapping works, such as~\cite{Doherty2019, gan2020bayesian}, as they are also using kernels to update the semantics as well}

Using the IPM and $\phi$, we can relate pixels in the segmented image to path coordinates. To represent the measurement likelihoods in all of $\Real_{\mathrm{SE}}^2$, we introduce an interpolation
\begin{equation}\label{eq:interpolation}
I^{\ell}_{\vvec} = \frac{\mathcal{K}(\|\vvec- \vvec_{\ell}\|_2)}{\sum_{\ell^{\prime}=1}^L \mathcal{K}(\|\vvec- \vvec_{\ell^{\prime}}\|_2)} \in[0,1].
\end{equation}
constructed with a kernel $\mathcal{K}\;:\;\Real^2_{\mathrm{SE}}\times \Real^2_{\mathrm{SE}}\mapsto \Real_{\geq 0}$. To capture locality and make the Bayesian updates computationally tractable~\cite{Doherty2019}, we use differentiable kernels with finite support, such as the sparse kernel in~\cite{melkumyan2009sparse}, with
\begin{align}
\mathcal{K}(d)  \hspace{-1.5pt}= \hspace{-1.5pt}
\sigma \left[\tfrac{2+\cos(2\pi \frac{d}{D})}{3}(1-\tfrac{d}{D})\hspace{-1.5pt}+\hspace{-1.5pt}\tfrac{1}{2\pi}\sin(2\pi \tfrac{d}{D}))\right],
\end{align}
if $d=\|\vvec - \vvec^{\ell}\| \geq D$, and 0 otherwise. Here, $D$ is the kernel bandwidth defining its support, and $\sigma>0$ is its amplitude.

\subsection{Categorical Updates with Semantic Information}
To model the semantic information obtained from image pixels projected into the 3D plane,  we use a categorical likelihood defined using the kernels in~\eqref{eq:interpolation} as
\begin{equation}\label{eq:interpolatedcategorical}
p(c|\Theta) = \prod_{\ell=1}^{L}\mathrm{C}(c; \wvec^{\ell})^{I^{\ell}_{\vvec}}.
\end{equation}
As the Dirichlet is the conjugate prior to the Categorical likelihood, the update with $L=1$ is well known, and by exponentiation with $I^{\ell}_{\vvec}$ we obtain a closed-form Bayesian update for the semantics, similar to that described in~\cite{gan2020bayesian}.

\begin{lemma}\label{lemma:dirspatial}
Given a prior in~\eqref{eq:priormodel} and a likelihood in~\eqref{eq:interpolatedcategorical}, from which a measurement $\ubar{c}\in\{1,...,K\}$ is sampled at a known point $\vvec\in\Real^d$, the posterior is $p(\Theta; \bar\Pcal)$, with parameters
\begin{subequations}\label{eq:dirupdate}
\begin{align}
\bar{a}_{i}^{\ell} &= a_{i}^{\ell} + I_{\vvec}^{\ell}[\ubar{c}=i],&
\forall&i\in[K], \ell\in[L],\\
(\bar\mu_i, \bar\lambda_i, \bar\alpha_i,\bar\beta_i) &= 
(\mu_i, \lambda_i, \alpha_i,\beta_i), &
\forall&i\in[K].
\end{align}
\end{subequations}
\end{lemma}

\begin{proof}
This follows by substituting~\eqref{eq:priormodel} and~\eqref{eq:interpolatedcategorical} in~\eqref{eq:bayes_update}, as the likelihood is independent of the normal-gamma variables.
\end{proof}

\subsection{Gaussian Mixture Updates with Properties}
\label{sec:momentmatching}
To model the road properties, such as friction parameters, we assume that these can be measured in the vehicle signals and that we can obtain instantaneous estimates as the car is driving over the road. We denote such a property measurement by $y$, and model the likelihood as a Gaussian mixture 
\begin{equation}\label{eq:extendedgmm}
p(y|\Theta) = \sum_{\ell=1}^{L}I^{\ell}_{\vvec} \sum_{i=1}^Kw_i^{\ell}\mathrm{N}(y; m_i, \tau_i^{-1}).
\end{equation}
Unlike the interpolated categorical, there is no conjugate prior for this measurement likelihood. The posterior computed by~\eqref{eq:bayes_update} can nonetheless be expressed in closed form.

\begin{lemma}\label{lemma:posterior:form}
Let $p(\Theta)$ in~\eqref{eq:priormodel} be the prior, with a measurement likelihood $p(y|\Theta)$ in~\eqref{eq:extendedgmm}. If $L=1$, the posterior given $\ubar{y}$ is
\begin{align}
p(\Theta|y)
= \frac{1}{M}\sum_{j=1}^K\Big(&u_j\mathrm{D}(\wvec|\avec_j^\star)c_j^\star\mathrm{N}\Gamma(m_j, \tau_j; \mu_j^\star, \lambda_j^\star, \alpha_j^\star,\beta_j^\star)\notag\\
&\prod_{i\neq j}\mathrm{N}\Gamma(m_i, \tau_i; \mu_i, \lambda_i, \alpha_i,\beta_i)\Big),
\end{align}
where
\begin{subequations}\label{eq:posteriorparameterexpr}
\begin{align}
\mu_j^\star &= \frac{\lambda_j\mu_j + \ubar{y}}{\lambda_j +1},&
\lambda_j^\star &= \lambda_j +1,\label{eq:muupdate}\\
\alpha_j^\star &= \alpha_j + 1/2,&
\beta_j^\star &= \beta_j + \lambda_j\frac{(\ubar{y} - \mu_j)^2}{2(1 + \lambda_j)},\label{eq:alphaupdate}\\
a_j^\star &= a_j + 1, & 
c_j^\star &= 
\frac{1}{\sqrt{2\pi}}\sqrt{\frac{\lambda_j}{\lambda_j^\star}}\frac{\Gamma(\alpha_j^*)}{\Gamma(\alpha_j)}\frac{(\beta_j)^{\alpha_i}}{(\beta_j^\star)^{\alpha_j^\star}},\label{eq:cstar}\\
M &= {\sum_{j= 1}^Ku_jc_j^\star}, &
u_j & =a_j \Big(\sum_{i=1}^K a_i\Big)^{-1}.
\end{align}
\end{subequations}
\end{lemma}

\begin{proof}
This follows along the lines of~\cite{jaini2016online} by a completion of squares to match the normal-gamma component.
\end{proof}

\begin{remark}
As $\alpha_j$ and $\beta_j$ generally increase with the number of incorporated measurements, it is necessary to evaluate $c_j^{\star}$ in the logarithm and explicitly use log-gamma functions (or approximate their ratio) for numerical stability.
\end{remark}

As $I_{\vvec}^{\ell}$ enters linearly, this can easily be extended to the case where $L>1$. Here,~\eqref{eq:muupdate} and~\eqref{eq:alphaupdate} remain the same, the weight update becomes similar to~\eqref{eq:dirupdate}. The number of terms in the posterior grows exponentially in the number of measurements included, but this can be remedied by BMM~\cite{jaini2016online}. The general idea of the BMM scheme is to project the posterior $p(\Theta|y)$ onto a distribution with the same parametric form as the prior. In our case, this would be a Dirichlet-Normal-Gamma in~\eqref{eq:priormodel}. We match the densities such that $\bar p(\Theta; \bar\Pcal)\simeq p(\Theta|y)$ in the sufficient moments. That is, we solve a system of equations $\mathbb{E}_{\bar p(\Theta; \bar\Pcal)}[g] =\mathbb{E}_{ p(\Theta|y)}[g]$ for the parameters $\bar\Pcal$, expressed in the moments $\mathbb{E}_{ p(\Theta|y)}[g]$.

\begin{lemma}\label{lemma:BMMexpressions}
\begin{subequations}The parameters of the matched density $\bar p(\Theta)$ are
\begin{align}
\bar\mu_i &= \mathbb{E}[m_i]&\forall&i\in[K],\\
\bar\lambda_i &=\frac{1}{\mathbb{E}[m_i^2\tau_i]-\mathbb{E}[m_i]^2\mathbb{E}[\tau_i]} &\forall&i\in[K],\\
\bar\alpha_i &= \frac{\mathbb{E}[\tau_i]^2}{\mathbb{E}[\tau_i^2] - \mathbb{E}[\tau_i]^2}&\forall&i\in[K],\\
\bar\beta_i &= \frac{\mathbb{E}[\tau_i]}{\mathbb{E}[\tau_i^2] - \mathbb{E}[\tau_i]^2}&\forall&i\in[K],\\
\bar a_i^\ell&= \mathbb{E}[w_i^\ell]\frac{\mathbb{E}[w_i^\ell]-\mathbb{E}[(w_i^\ell)^2]}{\mathbb{E}[(w_i^\ell)^2]-\mathbb{E}[w_i^\ell]^{2}}&\forall&i\in[K], \ell\in[L],
\end{align}
\end{subequations}
where the expectations are taken with respect to~\eqref{lemma:posterior:form}.
\end{lemma}
\begin{proof}
This is a simple extension of~\cite{jaini2016online}, and follows as all the factors of~\eqref{eq:priormodel} are modeled as independent.
\end{proof}

To compute the posterior given one property sample, $\ubar y$, we thus compute the parameters in~\eqref{eq:posteriorparameterexpr}, use these to express the moments $\mathbb{E}_{ p(\Theta|y)}[g]$ with $g \in\mathbb{S}$, and compute the matched posterior using Lemma~\ref{lemma:BMMexpressions}. In this case, the moments in Lemma~\ref{lemma:BMMexpressions} are known in closed form, and we can therefore efficiently incorporate the semantic measurements $\ubar{c}$ and property measurements $\ubar{y}$ in the same probabilistic map.

\subsection{Continuity Properties}

An appealing property of the considered measurement likelihoods is that some of their statistics adopt the same smoothness properties as the kernels used to model them. 

\begin{lemma}\label{lem:continuity}
Let $m_y(\vvec)$ and $V_y(\vvec)$ be the mean and variance of the posterior property likelihood. If the kernel defining $I_{\vvec}^{\ell}$ in~\eqref{eq:interpolation} is $n$-times continuously differentiable on its domain, where $k>0$, then $m_y\in\Ccal^n(\Real^{2}_{\mathrm{SE}}, \Real)$ and $V_y\in\Ccal^{n}(\Real^{2}_{\mathrm{SE}}, \Real)$.
\end{lemma}

\begin{proof}
By the chain rule, $n$-times continuously differentiability of the kernel implies the same property of $I^{\ell}_{\vvec}$ in $\vvec$. By marginalizing out $\Theta$, we have that $m_y(\vvec)=\sum_{\ell=1}^{L}I^{\ell}_{\vvec} \sum_{i=1}^KC_i^{\ell}$ for a set of constants $C_{i}^{\ell}$ that do not depend on $\vvec$, which then is $n$-times CD due to linearity in $I^{\ell}_{\vvec}$. A similar argument for $V_y(\vvec)$ concludes the proof.
\end{proof}

This statement holds irrespective of whether approximations are used to compute the moments, as $C_{i}^{\ell}$ is independent of $\vvec$. Furthermore, since inference is performed directly in path coordinates, we do not need the diffeomorphism when externalizing the posterior and relevant statistics for a controller. Consequently, the evaluation of these moments is fast and can be bounded in the support of the kernel and the density of the points $\{\vvec_{\ell}\}_{\ell=1}^L$, which are both design parameters.

\section{Numerical Results}
To demonstrate the utility of semantic property maps to vehicle control, we use data from a Lexus LC 500 for the transforms $\T(t)$ in sec~\ref{sec:map:ipm}. We create a true map $p(\Theta; \Pcal^{t})$ using known geometry of the road with $K=3$ classes, including gravel $(c=1)$, asphalt $(c=2)$, and water spots $(c=3)$, and generate $\Pcal^{t}$ such that the road has patches of gravel on the side of the road and several wet patches with their spatial locations randomized from a GP. We then estimate the map parameters $p(\Theta; \Pcal_{k})$, which, at a time-index $k$, has the parameters $\Pcal_{k}$. The prior $\Pcal_{0}$ is defined with parameters $\avec^{\ell} = (1,5,1)\;\forall \ell\in[L]$ and with properties differing from the true map by up to {$90$\%} in $\{(\mu_i,\lambda_i,\alpha_i,\beta_i)\}_{i=1}^K$.

To compare true and estimated maps, we use an average KL-divergence in the moments of the property likelihood
\begin{equation*}
\Lcal(\Pcal^t,\Pcal_k) = \frac1L\sum_{\ell = 1}^L\mathrm{KL}(\mathrm{N}(y; m^t_{\ell}, \Sigma^t_{\ell})\|\mathrm{N}(y; \hat m_{\ell,k}, \hat\Sigma_{\ell,k})),
\end{equation*}
where $(m^t_{\ell}, \Sigma^t_{\ell})$ and $(\hat m_{\ell,k}, \hat\Sigma_{\ell,k})$ is the mean and variance of the likelihood~\eqref{eq:extendedgmm} at $\vvec_{\ell}$ given $\Pcal^t$ and $\Pcal_k$, respectively.

\begin{figure}[t!]
    \centering
    \includegraphics[width=0.9\columnwidth, clip, trim={0 2cm 0 2cm}]{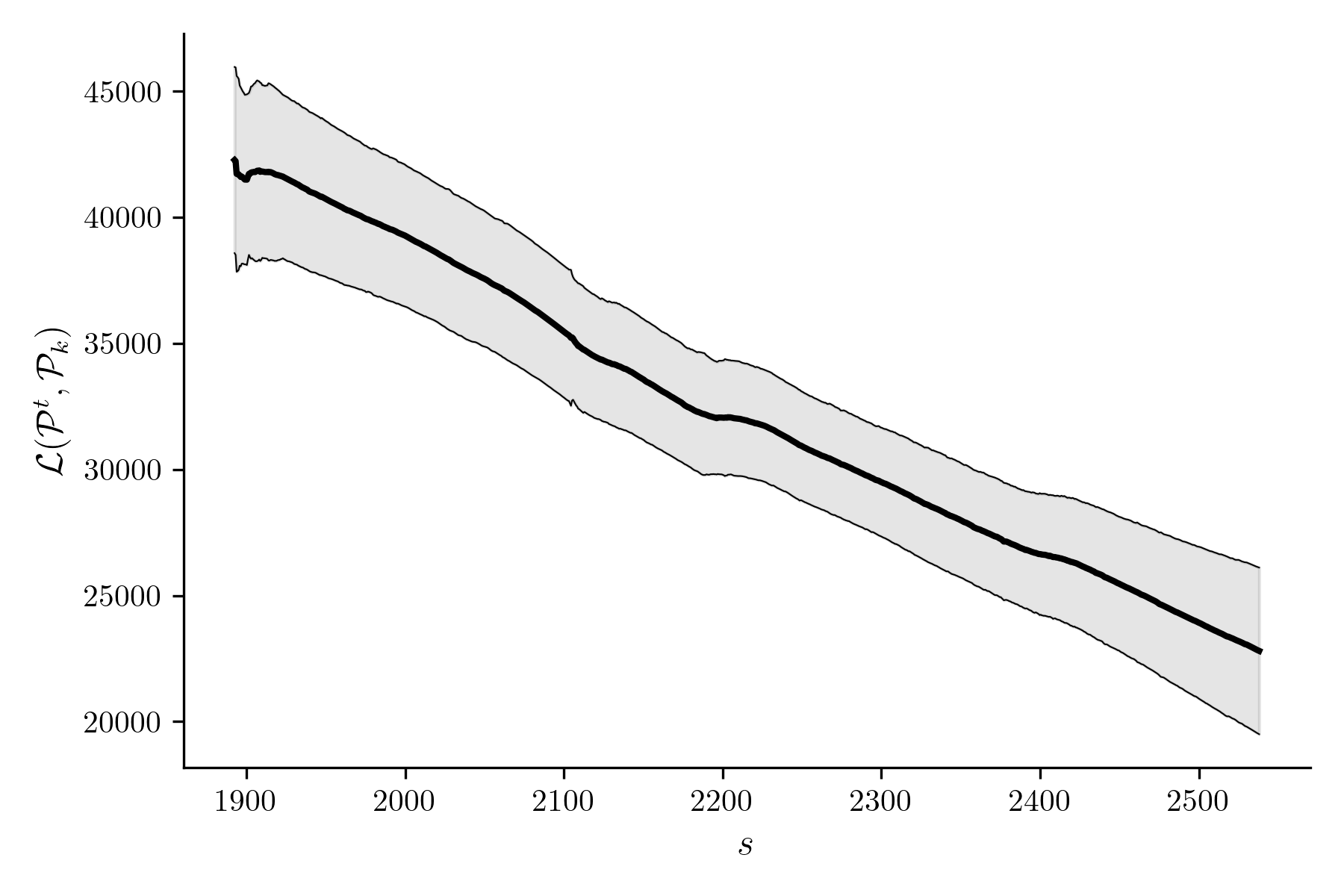}
    \vspace{-10pt}
    
    \caption{KL-divergence in the moments of the likelihood of the true map and the estimated map as a function of distance traveled, with a 95\% confidence interval.}
    \label{fig:likelihood}
\end{figure}

We start by realizing a set of $10$ true maps (one of which is shown in Fig.~\ref{fig:method}) and realizing semantics in the image space of a forward-facing camera at 20Hz, and friction estimates at 40Hz. The resulting estimates computed using Lemmas 1--4 are reported in Fig~\ref{fig:likelihood} over the first 600m of the map. Despite the large initial errors, the estimated maps approach the true maps as the vehicle drives along the road, with improvements over time in all tested map realizations.

\begin{figure}[t!]
    \centering
    \includegraphics[width=\columnwidth, clip, trim={0 0 0 10cm}]{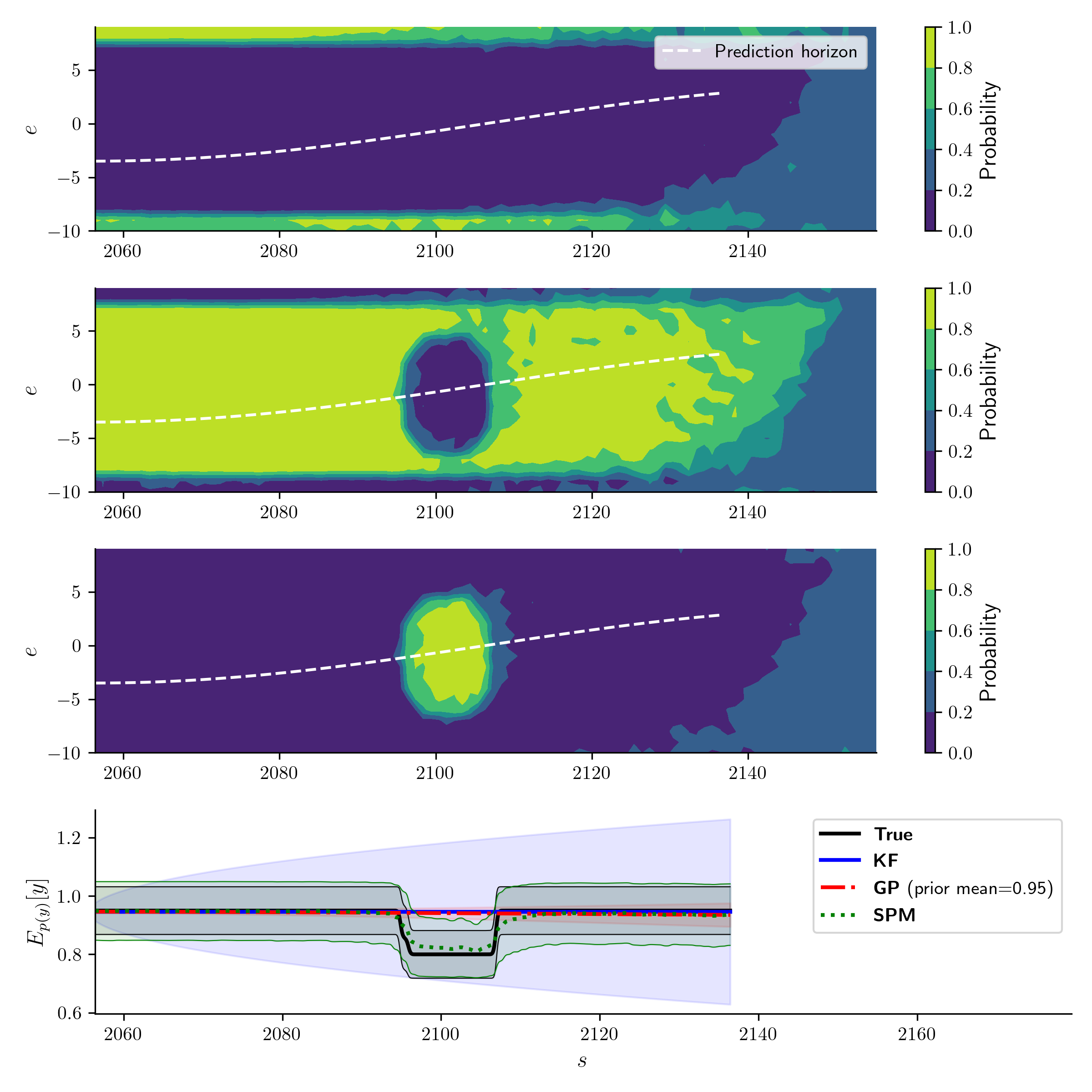}
    \vspace{-30pt}
    
    \caption{Difference in the property estimate error along a planning horizon in \emph{front of the vehicle}. \emph{Top:} Local class likelihoods ahead of the vehicle and a prediction horizon $\Vcal$, corresponding to one slice of the tensor in the externalization in Fig.~\ref{fig:method}. \emph{Bottom}: Predicted friction parameter along $\Vcal$.}
    \label{fig:prediction}
\end{figure}

\mg{Given that we are considering moments of the predictive likelihood, maybe it would be a good idea to evaluate a measure how close we are to the moments of the true likelihood? This would be a measure that we can study along trajectories in the context of Fig.~\ref{fig:prediction} as well. Written out mathematically, it would be something like
\begin{align}
    \mathcal{L}(\Pcal^t,\Pcal_k)= \int_{\Real_{\mathrm{SE}}^2}\mathrm{KL}(p(y; \Pcal^t)\|p(y; \Pcal_k))\delta(\vvec - \vvec^{\ell}) \der \vvec.
\end{align}
While we don't know $p(y; \Pcal)$ in closed form, we can approximate $p(y; \Pcal)\approx \sum_i w_i \mathrm{N}(y; \hat m_i, \hat \tau_i^{-1})$. However, even here there is no closed-form solution for the divergence (unless $K=1$). Upper and lower bounds exist, but it may be better to compute the moments of the likelihood directly. That is, we can compute
\begin{align*}
    \mathcal{L}(\Pcal^t,\Pcal_k) &= \int_{\Real_{\mathrm{SE}}^2}\mathrm{KL}(p(y; \Pcal^t)\|p(y; \Pcal_k))\delta(\vvec - \vvec^{\ell}) \der \vvec\\
    &\approx \int_{\Real_{\mathrm{SE}}^2}\mathrm{KL}(\hat{p}(y; \Pcal^t)\|\hat{p}(y; \Pcal_k))\delta(\vvec - \vvec^{\ell}) \der \vvec,
\end{align*}
where $\hat{p}(y; \Pcal) = \mathrm{N}(y | \mu_y(\vvec; \Pcal), V_y(\vvec; \Pcal))$. Doing so would allow us to directly compare the moments of the properties in the true map to those in the estimated map, and we could easily replace $\hat{p}(y; \Pcal_k)$ by a GP prediction.
}

Next, we demonstrate how the framework can benefit downstream control and planning applications by evaluating a vehicle's friction parameters along one prediction horizon of a controller. For simplicity, we let $\mathcal{D} = \{(y, s, e)\in\Real\times \Real^2_{\mathrm{SE}}|s_0-100 < s < s_0\}$ denote all property measurements sampled in the past 100m, and let $\mathcal{V} = \{(s, e)\in\Real^2_{\mathrm{SE}}|s=s_0+n, e=3.5\cos(\tfrac{\pi}{10}(s-s_0), n\in[N]\}$. For this evaluation, we let $N=80$ and consider three methods:
\begin{itemize}
    \item \textbf{KF}: A Kalman Filter (KF), with the property as its state and a small random walk as its dynamics. The estimate along $\Vcal$ is computed by prediction from the marginal filtering posterior at $s_0$. This is analogous to parameter estimation methods in adaptive MPC~\cite{thompson2024adaptive}.
    \item \textbf{GP}: Training a GP on $\mathcal{D}$ with RBF kernels whose hyperparameters are found by minimizing the marginal likelihood online. The mean prediction is taken over $\Vcal$, and the result resembles techniques commonly used in data-driven predictive control~\cite{berntorp2021joint}.
    \item \textbf{SPM}: The semantic property map proposed in this work, updated both with semantics and property measurements, evaluated along the horizon $\Vcal$.
\end{itemize}

The result is shown in Fig.~\ref{fig:prediction}, and demonstrates that the \textbf{KF} and \textbf{GP} yield substantial prediction errors when there is variability in the property over $\Vcal$. On the other hand, the proposed \textbf{SPM} captures this variability by leveraging the semantic information, providing improved predictions. It is therefore likely to affect the controller or planner that uses these predictions. Future work will explore the impact of this improved property prediction on controller performance.

\section{Conclusions}
In this paper, we present a framework for estimating the parameters of a vehicle in driving applications. Rather than relying on single-modality vehicle signals, we fuse semantic information from cameras with proprioceptive sensing in a probabilistic map, and demonstrate the efficacy of using Bayesian moment matching for this purpose. The map, constructed with conjugate priors and defined in the same path coordinates as the vehicle controller, captures spatial variations in parameter likelihoods. This results in an online-adapted map that enhances the control system with smooth parameter likelihood moments. Extensions of this work to multivariate property distributions and integration with real vehicle controllers will be considered in future work.

\bibliographystyle{IEEEbib}
\bibliography{references}

@article{margolis2023learning,
  title={Learning to see physical properties with active sensing motor policies},
  author={Margolis, Gabriel B and Fu, Xiang and Ji, Yandong and Agrawal, Pulkit},
  journal={arXiv preprint arXiv:2311.01405},
  year={2023}
}

@article{lew2024risk,
  title={Risk-Averse Model Predictive Control for Racing in Adverse Conditions},
  author={Lew, Thomas and Greiff, Marcus and Djeumou, Franck and Suminaka, Makoto and Thompson, Michael and Subosits, John},
  journal={Int. Conf. on Robotics and Automation (ICRA)},
  year={2025},
  publisher={IEEE}
}

@inproceedings{mccormac2017semanticfusion,
  title={Semanticfusion: Dense 3d semantic mapping with convolutional neural networks},
  author={McCormac, John and Handa, Ankur and Davison, Andrew and Leutenegger, Stefan},
  booktitle={Int. Conf. on Robotics and Aut.},
  pages={4628--4635},
  year={2017},
  organization={IEEE}
}

@inproceedings{zhou2024feature,
    title={Feature 3dgs: Supercharging 3d gaussian splatting to enable distilled feature fields},
    author={Zhou, Shijie and Chang, Haoran and Jiang, Sicheng and Fan, Zhiwen and Zhu, Zehao and Xu, Dejia and Chari, Pradyumna and You, Suya and Wang, Zhangyang and Kadambi, Achuta},
    booktitle={Conf. on Computer Vision and Pattern Recognition (CVPR)},
    pages={21676--21685},
    year={2024}
}

@article{o2012gaussian,
  title={Gaussian process occupancy maps},
  author={O’Callaghan, Simon T and Ramos, Fabio T},
  journal={The International Journal of Robotics Research},
  volume={31},
  number={1},
  pages={42--62},
  year={2012},
  publisher={SAGE Publications Sage UK: London, England}
}

@article{Doherty2019,
  doi = {10.1109/tro.2019.2912487},
  url = {https://doi.org/10.1109/tro.2019.2912487},
  year = {2019},
  publisher = {Institute of Electrical and Electronics Engineers ({IEEE})},
  pages = {1--14},
  author = {Kevin Doherty and Tixiao Shan and Jinkun Wang and Brendan Englot},
  title = {Learning-Aided 3-D Occupancy Mapping With Bayesian Generalized Kernel Inference},
  journal = {{IEEE} Transactions on Robotics}
}

@article{gan2020bayesian,
  title={Bayesian spatial kernel smoothing for scalable dense semantic mapping},
  author={Gan, Lu and Zhang, Ray and Grizzle, Jessy W and Eustice, Ryan M and Ghaffari, Maani},
  journal={Robotics and Automation Letters},
  volume={5},
  number={2},
  pages={790--797},
  year={2020},
  publisher={IEEE}
}

@article{wilson2025latentbki,
  title={LatentBKI: Open-Dictionary Continuous Mapping in Visual-Language Latent Spaces With quantifiable uncertainty},
  author={Wilson, Joey and Xu, Ruihan and Sun, Yile and Ewen, Parker and Zhu, Minghan and Barton, Kira and Ghaffari, Maani},
  journal={IEEE Robotics and Automation Letters},
  year={2025},
  publisher={IEEE}
}

@article{wilson2024convbki,
  title={Convbki: Real-time probabilistic semantic mapping network with quantifiable uncertainty},
  author={Wilson, Joey and Fu, Yuewei and Friesen, Joshua and Ewen, Parker and Capodieci, Andrew and Jayakumar, Paramsothy and Barton, Kira and Ghaffari, Maani},
  journal={IEEE Transactions on Robotics},
  year={2024},
  publisher={IEEE}
}

@article{jaini2016online,
  title={Online and distributed learning of Gaussian mixture models by Bayesian moment matching},
  author={Jaini, Priyank and Poupart, Pascal},
  journal={arXiv preprint arXiv:1609.05881},
  year={2016}
}

@article{mukhtar2015vehicle,
  title={Vehicle detection techniques for collision avoidance systems: A review},
  author={Mukhtar, Amir and Xia, Likun and Tang, Tong Boon},
  journal={IEEE Trans. on Int. Trans. Sys.},
  volume={16},
  number={5},
  pages={2318--2338},
  year={2015},
  publisher={IEEE}
}

@article{Goh2019,
  title={Toward automated vehicle control beyond the stability limits: drifting along a general path},
  author={Goh, Jonathan Y and Goel, Tushar and Christian Gerdes, J},
  journal={Journal of Dynamic Systems, Measurement, and Control},
  volume={142},
  number={2},
  pages={021004},
  year={2020},
  publisher={American Society of Mechanical Engineers}
}

@article{nuchter2008towards,
  title={Towards semantic maps for mobile robots},
  author={N{\"u}chter, Andreas and Hertzberg, Joachim},
  journal={Robotics and Autonomous Systems},
  volume={56},
  number={11},
  pages={915--926},
  year={2008},
  publisher={Elsevier}
}

@inproceedings{melkumyan2009sparse,
  title={A sparse covariance function for exact Gaussian process inference in large datasets.},
  author={Melkumyan, Arman and Ramos, Fabio},
  booktitle={IJCAI},
  volume={9},
  pages={1936--1942},
  year={2009}
}

@article{ewen2022these,
  title={These maps are made for walking: Real-time terrain property estimation for mobile robots},
  author={Ewen, Parker and Li, Adam and Chen, Yuxin and Hong, Steven and Vasudevan, Ram},
  journal={Rob. and Aut. Lett.},
  volume={7},
  number={3},
  year={2022},
  publisher={IEEE}
}

@article{ewen2024you,
  title={You've Got to Feel It To Believe It: Multi-Modal Bayesian Inference for Semantic and Property Prediction},
  author={Ewen, Parker and Chen, Hao and Chen, Yuzhen and Li, Anran and Bagali, Anup and Gunjal, Gitesh and Vasudevan, Ram},
  journal={arXiv preprint arXiv:2402.05872},
  year={2024}
}

@article{gustafsson1997slip,
  title={Slip-based tire-road friction estimation},
  author={Gustafsson, Fredrik},
  journal={Automatica},
  pages={1087--1099},
  year={1997},
  publisher={Elsevier}
}

@article{dallas2020online,
  title={Online terrain estimation for autonomous vehicles on deformable terrains},
  author={Dallas, James and Jain, Kshitij and Dong, Zheng and Sapronov, Leonid and Cole, Michael P and Jayakumar, Paramsothy and Ersal, Tulga},
  journal={Journal of Terramechanics},
  volume={91},
  pages={11--22},
  year={2020},
  publisher={Elsevier}
}

@ARTICLE{du20,
  author={Du, Yuchuan and Liu, Chenglong and Song, Yang and Li, Yishun and Shen, Yu},
  journal={Trans. on Intell. Transportation Systems}, 
  title={Rapid Estimation of Road Friction for Anti-Skid Autonomous Driving}, 
  year={2020},
  volume={21},
  number={6},
  pages={2461-2470},
  doi={10.1109/TITS.2019.2918567}}

@INPROCEEDINGS{panahandeh17,
  author={Panahandeh, Ghazaleh and Ek, Erik and Mohammadiha, Nasser},
  booktitle={2017 IEEE Intelligent Vehicles Symposium}, 
  title={Road friction estimation for connected vehicles using supervised machine learning}, 
  year={2017},
  volume={},
  number={},
  pages={1262-1267},
  doi={10.1109/IVS.2017.7995885}}

@article{thompson2024adaptive,
  title={Adaptive nonlinear model predictive control: maximizing tire force and obstacle avoidance in autonomous vehicles},
  author={Thompson, Michael and Dallas, James and Goh, Jonathan YM and Balachandran, Avinash},
  journal={Trans. on Field Robotics},
  year={2024},
  publisher={IEEE}
}

@article{berntorp2024framework,
title = {A framework for joint vehicle localization and road mapping using onboard sensors},
journal = {Control Engineering Practice},
volume = {153},
pages = {106112},
year = {2024},
issn = {0967-0661},
doi = {https://doi.org/10.1016/j.conengprac.2024.106112},
url = {https://www.sciencedirect.com/science/article/pii/S0967066124002715},
author = {Karl Berntorp and Marcus Greiff},
}

@inproceedings{berntorp2021joint,
  title={Joint Tire-Stiffness and Vehicle-Inertial Parameter Estimation for Improved Predictive Control},
  author={Berntorp, Karl and Quirynen, Rien and Vaskov, Sean},
  booktitle={2021 American Control Conference (ACC)},
  pages={186--191},
  year={2021},
  organization={IEEE}
}

@article{svensson2021traction,
  title={Traction adaptive motion planning and control at the limits of handling},
  author={Svensson, Lars and Bujarbaruah, Monimoy and Karsolia, Arpit and Berger, Christian and T{\"o}rngren, Martin},
  journal={Trans. on Cont. Sys. Tech.},
  volume={30},
  number={5},
  pages={1888--1904},
  year={2021},
  publisher={IEEE}
}

@inproceedings{svensson2021fusion,
  title={Fusion of heterogeneous friction estimates for traction adaptive motion planning and control},
  author={Svensson, Lars and T{\"o}rngren, Martin},
  booktitle={Int. Intell. Trans. Sys. Conf.},
  pages={424--431},
  year={2021},
  organization={IEEE}
}

@article{werner2025gripmap,
  title={GripMap: An Efficient, Spatially Resolved Constraint Framework for Offline and Online Trajectory Planning in Autonomous Racing},
  author={Werner, Frederik and Schwehn, Ann-Kathrin and Lienkamp, Markus and Betz, Johannes},
  journal={arXiv preprint arXiv:2504.12115},
  year={2025}
}

\end{document}